\documentclass[reprint,twocolumn,aps,prb,superscriptaddress]{revtex4-2}
\usepackage[T1]{fontenc}
\usepackage[utf8]{inputenc}
\usepackage{lmodern}
\usepackage{graphicx}
\usepackage{amsmath}
\usepackage{xcolor}
\usepackage{float}
\usepackage{hyperref}
\usepackage{braket}
\usepackage{amssymb,amsmath,amsthm, array}
\usepackage{mdframed}
\usepackage{mathtools}
\usepackage{yfonts}
\usepackage{comment}

\DeclareUnicodeCharacter{0301}{-}

\begin{document}
\renewcommand{\vec}[1]{\mathbf{#1}}
\newcommand{\ii}{\mathrm{i}}
\def\ya#1{{\color{orange}{#1}}}

\def\d{\downarrow}
\def\u{\uparrow}
\def\nn{\nonumber}

\title{Majorana Signatures in the Tripartite Uncertainty Relations with Quantum Memory}

\author{D. Maroulakos}
\affiliation{Doctoral School at the  University of Rzesz\'ow, 35-317 Rzesz\'ow, Poland}
\author{C. Jasiukiewicz}
\affiliation{Department of Physics and Medical Engineering, Rzesz\'ow University of Technology, 35-959 Rzesz\'ow, Poland}
\author{A. Wal}
\affiliation{Institute of Physics, Faculty of Exact and Technical Sciences, University of Rzesz\'ow, 35-317 Rzesz\'ow, Poland}
\author{A. Sinner}
\affiliation{Institute of Physics, University of Opole, 45-052 Opole, Poland}
\author{I. Weymann}
\affiliation{Institute of Spintronics and Quantum Information,
Faculty of Physics and Astronomy, A. Mickiewicz University, 61-614 Pozna\'n, Poland}
\author{T. Domański}
\affiliation{Institute of Physics, M. Curie-Sk\l{}odowska University, 20-031 Lublin, Poland}
\author{L. Chotorlishvili}
\affiliation{Department of Physics and Medical Engineering, Rzesz\'ow University of Technology, 35-959 Rzesz\'ow, Poland}

\date{\today}
\begin{abstract}
Quantumness imposes a fundamental limit on measurement accuracy.
The paradigmatic cases are Heisenberg's uncertainty relation in the original formulation,
Robertson's formulation, and improved uncertainty relations.
However, the more universal measures are given in terms of quantum entropies.
Uncertainties of measurements done on one quantum system
correlated with another quantum system constitute a more intriguing question.
Quantum correlations can influence the lower bound of uncertainties,
and the reason for this is the quantum memory.
In this article, we study uncertainties of measurements
performed on one quantum dot correlated with the second one through the superconductor, hosting the Majorana boundary modes.
We prove that the Majorana quasiparticles allow the uncertainties to reach the minimal possible lower bound.
By rigorous theoretical considerations, we obtain the result of experimental relevance expressed in terms of only two parameters: the overlap between Majorana modes and their coupling strength with the quantum dots. We show that the overlap between Majorana modes reduces quantum uncertainties, which is a general result of fundamental importance. We also propose the protocol to measure spins in both quantum dots, consecutively, and demonstrate that the result of the second measurement would depend on the presence of Majorana quasiparticles. This could serve as an indirect tool for their empirical observation, which is of importance for the ongoing discussions concerning unambiguous detection of the Majorana quasiparticles in nanoscopic hybrid structures.

\end{abstract}

\maketitle

\section{Introduction}
\label{sec:Introduction}
If two Hermitian operators $\hat a$ and $\hat b$ do not commute $[\hat a,\hat b]\neq 0$,
the uncertainty principle asserts the fundamental limit
on the accuracy of consecutive measurements done on $\hat a$ and $\hat b$.
A prominent example is the Heisenberg's uncertainty principle,
which says that due to the non-commutativity of momentum
and coordinate operators  $\left[\hat p,\hat x\right]\neq 0 $, pin-point measurement
of the coordinate operator $\hat x$ reduces the accuracy
of measuring the momentum operator $\hat p$,
and vice versa, i.e., $\Delta x\Delta p\geqslant \hbar$, where $\hbar$ is the Planck's constant.
Following Robertson, one may write uncertainty relations in the form \cite{RevModPhys.89.015002}: 
\begin{eqnarray}\label{Robertson}
\Delta a\cdot\Delta b\geqslant \frac{1}{2}\left\vert\bra{\psi}[\hat a,\hat b]\ket{\psi}\right\vert, 
\end{eqnarray}
where $\Delta O=\sqrt{\langle \hat O^2\rangle-\langle\hat O\rangle^2}$. 
While Eq.~(\ref{Robertson}) looks more general than the Heisenberg's uncertainty relation,
when  $\ket{\psi}$ is the eigenfunction of  $\hat a$ or $\hat b$, Eq.~(\ref{Robertson}) takes a trivial form.
Therefore, one has to utilize more advanced entropic uncertainty relations
to eliminate arbitrariness of the choice of state  $\ket{\psi}$.
The right-hand side of inequality Eq.~(\ref{Robertson}) defines the minimal possible measurement uncertainty.
In contrast, the left-hand side quantifies the uncertainty
of actual measurements and can exceed the right-hand side.
The measured system $A$ might be correlated with another quantum system $B$, forming an entangled state $\hat\rho_{AB}$.
Correlations between the two systems influence
the uncertainty of quantum measurements performed on the system $A$.
In the seminal work \cite{berta2010uncertainty} it has been shown that quantum memory can affect the bound of entropic uncertainty relations. Different aspects of quantum memory have been addressed in Refs.~\cite{wang2019quantum,ming2020improved,dolatkhah2020tightening,bergh2021entanglement, PhysRevA.104.062204,chotorlishvili2019spin,song2022environment,zhu2021zero,kurashvili2022quantum, PhysRevD.103.036011,kurashvili2022quantum}.

\begin{figure}[t]
\centerline{\includegraphics[width=0.5\textwidth]{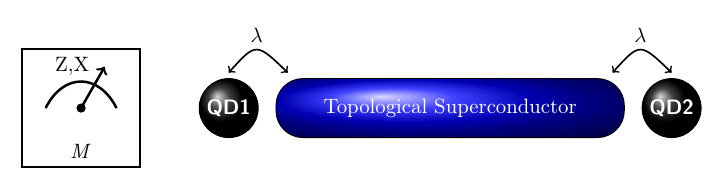}}
\caption{The tripartite system, consisting of two quantum dots, QD$_1$={\bf Alice} and QD$_2$={\bf Bob},
	and a topological superconductor={\bf Charlie}, hosting the Majorana modes.
	The two noncommuting operators from QD$_1$: $\hat\sigma_x$ and $\hat\sigma_z$ are measured.
	According to the improved entropic uncertainty relation (EUR),
	the lower bound of the uncertainty of those measurements
	depends on the correlations between subsystems: \textbf{[Alice, Charlie]} and \textbf{[Alice, Bob]}.}
\label{fig1}
\end{figure}

In this article we carry out rigorous studies of the quantum memory, which can in principle be stored in a device comprising two quantum dots connected by a short topological wire with the boundary Majorana modes, Fig.\ \ref{fig1}. It has been recently shown \cite{Cayao-2024,Majoranas1-2025} that in such system the degree of entanglement between the quantum dots can be varied by tuning the coupling strength with both Majorana modes. As it turns out, this is only one particular aspect of the larger picture we have unveiled in the present study. In a surprisingly unexpected manner, the interference between Majorana modes emerges as a powerful tool for controlling the quantumness of the whole device. We predict that the interference between Majorana modes dramatically influences fundamental uncertainties of quantum measurements. Namely, it reduces  the quantum uncertainties of experimentally feasible spin measurements in quantum dots.

Recent intensive studies of the Majorana fermions explored variety of platforms, mostly one- and two-dimensional magnetic samples in contact with bulk superconductors \cite{beenakker2013search,RevModPhys.87.137,PhysRevB.84.144522,review-2020,review-2021,Wiesendanger-2025,Zou2023}. Moreover, the investigations have encompassed, among others, Majorana quasiparticles in the presence of dissipation by environmental modes \cite{PhysRevLett.125.147701}, using the Born-Markov approximation and Lindblad formalism \cite{PhysRevB.102.134501}. There have also been studies on finite temperature effects and thermoelectric fluctuations of interfering Majorana bound states \cite{PhysRevB.111.125402}. Nonequilibrium quantum noise has been considered as a possible tool for probing the Majorana bound states via the fluctuation fingerprints \cite{PhysRevB.109.195410}. Furthermore, shot noise and conductance have been investigated in the context of nonequilibrium states  \cite{PhysRevB.105.205430}. The Majorana modes have been studied in the field of quantum thermodynamics as well, considering a two-terminal quantum spin-Hall heat engine and refrigerator with embedded Majorana bound states  \cite{mishra2024majorana}. In fact, there are plenty of other interesting physical aspects connected with Majorana physics, such as the Aharonov-Bohm oscillations \cite{zou2022aharonov}, Majorana ensembles with fractional entropy and entropy revival via tunneling phases \cite{PhysRevB.104.205406,PhysRevB.103.075440}, to name a few. In what follows, we focus on a different yet equally interesting aspect of such hybrid systems. In particular, we analyze tripartite quantum entanglement of quantum dots attached to the topological superconductor with the Majorana boundary modes.

\section{Quantum-memory-assisted entropic uncertainties}
\label{sec:uncertainties}

The original quantum-memory-assisted entropic uncertainty relation (EUR)
was formulated for the bipartite quantum system $\hat\varrho_{AB}$ shared between {\bf Alice} $A$ and {\bf Bob} $B$
\cite{berta2010uncertainty,wilde2013quantum,PhysRevB.108.134411}.
{\bf Alice} performs two positive operator-valued measurements (POVMs)
on the \textbf{A} qubit at her hand. {\bf Alice} uses projectors 
$\Pi_{1,2}^{zA}=\ket{\psi_{1,2}}\bra{\psi_{1,2}}$, $\Pi_{1,2}^{xA}=\ket{\phi_{1,2}}\bra{\phi_{1,2}}$, where $\ket{\psi_{1,2}}\equiv \ket{0}_A,\,\ket{1}_A$ and $\ket{\phi_{1,2}}=\frac{1}{\sqrt{2}}\left(\ket{0}_A\pm\ket{1}_A\right)$ are eigenfunctions of $z$ and $x$ components of the qubit $\textbf{A}$. Then EUR for the bipartite system is given by
\begin{eqnarray}
\label{eq:BobsIgnorance}
S(X\vert B)+S(Z\vert B)\geqslant \log_2\frac{1}{c}+S(A\vert B),
\end{eqnarray}
where $c=\max\left\lbrace \vert\bra{\psi_i}\ket{\phi_j}\vert^2\right\rbrace$
is a measure of complimentarity and $S(X\vert B)=S(\hat\rho^X_{AB})-S(\hat\rho_B)$, $S(Z\vert B)=S(\hat\rho^Z_{AB})-S(\hat\rho_B)$, with $S(\hat\varrho)=-\hat\varrho\log_2\hat\varrho$
being conditional quantum entropies of the states \cite{wilde2013quantum}
\begin{eqnarray}
\label{eq:rhoZ}
&&\hat\rho^Z_{AB}=\sum\limits_{n=1,2}\Pi_{n}^{z}\otimes {\text{Tr}}_A\left\lbrace(\Pi_{n}^{z}\otimes\hat I_B)\hat\rho_{AB}\right\rbrace,\\
\label{eq:rhoX}
&&\hat\rho^X_{AB}=\sum\limits_{n=1,2}\Pi_{n}^{x}\otimes {\text{Tr}}_A\left\lbrace(\Pi_{n}^{x}\otimes\hat I_B)\hat\rho_{AB}\right\rbrace.
\end{eqnarray}
We note that the left-hand side of Eq.~(\ref{eq:BobsIgnorance}) defines the uncertainty about measurement results (two measurements done on the $\textbf{x}$ and $\textbf{z}$ spin components of the system \textbf{A}). The right-hand side of Eq.(\ref{eq:BobsIgnorance}) defines the lower bound of this uncertainty. The first term on the right-hand side $\log_2(1/c)$ is positive. However, conditional quantum entropy $S(A\vert B)$ can be negative for entangled states, and when it is negative,  it reduces the lower bound of uncertainty. Negative conditional quantum entropy $S(A\vert B)$ means that the state $\hat\varrho_{AB}$ for sure is entangled, but the converse is not always true. The quantum memory defines the bound of minimal theoretically possible uncertainty given by $\max\left\lbrace \vert\bra{\psi_i}\ket{\phi_j}\vert^2\right\rbrace$.

\section{The setup}
\label{sec: setup}
The system of our interest here is not bipartite but tripartite (Fig.\ \ref{fig1}). Specifically, we consider two quantum dots (QD$_1$, QD$_2$) indirectly contacted via the Majorana modes, existing at boundaries of the topologically nontrivial superconducting nanowire \cite{kitaev2001unpaired}. Majorana quasiparticles do always exist in pairs and obey non-Abelian statistical rules, that would be appealing for implementation of quantum computations \cite{PhysRevLett.104.040502,alicea2012new}. However, despite extensive studies \cite{beenakker2013search,RevModPhys.87.137,PhysRevB.84.144522,review-2020,review-2021,Wiesendanger-2025}, it is not yet clear whether they are cross-correlated or not \cite{teleportation-2016,teleportation-2024}. 
Suitable platforms to address this issue are hybrid structures with side-attached quantum dots \cite{PhysRevB.84.140501,PhysRevB.84.201308,Deng-2018,PhysRevB.109.075432},
where QD$_{i}$ can be regarded as detectors of nonlocal correlations \cite{PhysRevB.110.035413}.
The low-energy model of our hybrid structure can be described by the following Hamiltonian:
\begin{eqnarray}\label{Total Hamiltonianone}
\hat H=\sum\limits_{i=1,2}\hat H^{QD}_{i}+\hat V.
\end{eqnarray}
Quantum dots are treated as the Anderson type impurities 
$\hat H^{QD}_{i}=\sum\limits_\sigma\varepsilon_{i\sigma} d^\dag_{i\sigma}d_{i\sigma}+U\hat n_{i\uparrow}\hat n_{i\downarrow}$,
where $d^\dag_{i\sigma},\,d_{i\sigma}$ are creation and annihilation operators of the electrons with spin $\sigma=\uparrow,\,\downarrow$ and energy $\varepsilon_{i\sigma}$,
and $U$ denotes the strength of the Coulomb repulsion between opposite spin electrons,
while $\hat n_{i\sigma}=d^\dag_{i\sigma}d_{i\sigma}$. 
QDs are hybridized with the Majorana quasiparticles via
\begin{eqnarray}
\hat V=\lambda_1\left(d^\dag_{1\uparrow}-d_{1\uparrow}\right)\hat\gamma_1+\nonumber
i\lambda_2\hat\gamma_2\left(d^\dag_{2\uparrow}+d_{2\uparrow}\right)+i\varepsilon_M\hat\gamma_1\hat\gamma_2 .
\label{Total Hamiltonianonethree}
\end{eqnarray}
We assume that only $\uparrow$-spin electrons of the quantum dots are coupled to these Majorana boundary modes with the coupling strength $\lambda_i$. The overlap $\varepsilon_M$ between the Majorana modes is finite when the topological superconducting nanowire is shorter than the superconducting coherence length. Operators $\hat\gamma_i$ and $\hat\gamma^\dag_i$ are self-hermitian, $\hat{\gamma}_{i}^{\dagger}=\hat{\gamma}_{i}$. It is convenient to express them in terms of conventional fermion operators defined by
$\hat\gamma_1=(\hat f^\dag+\hat f)/\sqrt{2}$, $\hat\gamma_2=i(\hat f^\dag-\hat f)/\sqrt{2}$. 
We introduce the occupancy representation $|n^{}_f,n^{}_{d^{}_1},n^{}_{d^{}_2}\rangle$, enumerating the basis vectors as follows
$\ket{\varphi_1}=\ket{0,0,0}$, $\ket{\varphi_2}=\ket{1,0,0}$, $\ket{\varphi_3}=\ket{0,1,0}$, $\ket{\varphi_4}=\ket{1,1,0}$, $\ket{\varphi_5}=\ket{0,0,1}$, 
$\ket{\varphi_6}=\ket{1,0,1}$, $\ket{\varphi_7}=\ket{0,1,1}$, $\ket{\varphi_8}=\ket{1,1,1}$. 
Technical details can be found in Ref.~\cite{Majoranas1-2025}.
For the sake of analytic study we consider the parameter set:
$\varepsilon^{}_{1\u,\d}=0$, $\varepsilon^{}_{2\u,\d}=0$, $U=0$, $\varepsilon_M = 2\omega$, $\lambda^{}_1=-\lambda^{}_2=\sqrt{2}\lambda$. Then, the energy eigenbasis of the system is given by:
$\ket{e_1}=-\eta_+\ket{1_f}\otimes\ket{\Phi_d^-}+\xi_+\ket{0_f}\otimes\ket{\Psi_d^+}$,
$\ket{e_2}=\eta_-\ket{1_f}\otimes\ket{\Phi_d^-}+\xi_-\ket{0_f}\otimes\ket{\Psi_d^+}$,
 $E_1=-\sqrt{\omega^2+4\lambda^2}=-\Delta$,
 $\Delta = \sqrt{\omega^2 + 4\lambda^2}$, 
  $\ket{e_3}=-\ket{0_f}\otimes\ket{\Psi_d^-}$, 
$\ket{e_4}=-\ket{0_f}\otimes\ket{\Phi_d^-}$, $E_2=-\omega$; $\ket{e_5}=\ket{1_f}\otimes\ket{\Phi_d^+}$, $\ket{e_6}=\ket{1_f}\otimes\ket{\Psi_d^+}$, $E_3=\omega$;
$\ket{e_7}=-\eta_-\ket{1_f}\otimes\ket{\Phi_d^-}+\xi_-\ket{0_f}\otimes\ket{\Psi_d^+}$,
$\ket{e_8}=\eta_+\ket{0_f}\otimes\ket{\Phi_d^+}+\xi_+\ket{1_f}\otimes\ket{\Psi_d^-}$,
$E_4=\Delta$, where we introduced the notation: $\eta_\pm=\frac{2\lambda}{\sqrt{4\lambda^2+(\omega\pm\Delta)^2}}$, $\xi_\pm=\frac{\omega\pm\Delta}{\sqrt{4\lambda^2+(\omega\pm\Delta)^2}}$ and $\ket{\Psi^\pm_d}$, $\ket{\Phi^\pm_d}$ are Bell states of the quantum dots.

\section{Indirect observation of Majorana modes}
\label{sec: Indirect observation}

Quantum entanglement of the hybrid setup displayed in Fig.\ \ref{fig1} has been a topic of recent interest \cite{Cayao-2024,Majoranas1-2025} and its empirical realization seems to be feasible. For example, the density matrices of tri- and bipartite states of the hybrid QD$_{1}$ - topological nanowire - QD$_{2}$ system can be constructed by the quantum state tomography method \cite{steffen2006measurement}. Regarding a finite overlap between the Majorana modes, $\varepsilon_{M} \neq 0$, such a situation could be encountered e.g.\ in short-length nanowires deposited on superconducting substrates \cite{Wiesendanger-2025} or in the minimal Kitaev chain, consisting of a few semiconducting quantum dots interconnected through the conventional superconductor \cite{dvir2023realization}. A single-shot measurement scheme would be capable to selectively access the singlet or triplet pairing states induced in the quantum dots and could mimic the POVMs discussed above \cite{PhysRevB.74.195303}. 

Before addressing the quantum uncertainty, we would like to emphasize that indirect information about the Majorana modes can be inferred from the following measurement protocol: (a) by first measuring the spin of the first QD and then the spin of the second QD; (b) by measuring the spin of second QD directly. 
Our focus here are the Quantum Witnesses (QW), which quantify the invasiveness of quantum measurements \cite{PhysRevA.92.032101,li2012witnessing, PhysRevLett.122.070603, PhysRevA.87.052115, PhysRevD.103.036011}. 
Due to the coupling between QDs through the Majorana modes, the results of measurements of the second spin in the two cases (a) and (b) should be different. Such difference in the measurement results could hence provide evidence for an indirect observation of Majorana modes. 

We consider the most general case and assume that the quantum system between measurements evolves through the trace-preserving map $\mathcal{\hat{F}}$, which is not necessarily a unitary operation. The projector measurement operators for qubits read ${\Pi^{\mathrm{s}}_{j\alpha}=|\alpha\rangle\langle\alpha|}$. Here, the index $j$ defines the subsystems of QDs $j=A, B$ and $\alpha=0,1$ refers to the ground and excited spin states. The direct measurement probability is given by $\mathcal{P}^{j\alpha}_{\hat{\varrho}}=\mathrm{Tr}\{\Pi^{\mathrm{s}}_{j\alpha}\mathcal{\hat{F}}[\hat{\varrho}]\}$.
On the other hand, the indirect measurements probability reads ${\mathcal{Q}^{j\alpha}_{\hat{\varrho}}=\mathrm{Tr}\{\Pi^{\mathrm{s}}_{j\alpha}\mathcal{\hat{F}}[\hat{\varrho}_{\mathrm{post}}]\}}$, where $\hat{\varrho}_{\mathrm{post}}$ is the post-measurement state formed after the measurement in the indirect measurement scheme. States evolved through the trace-preserving maps are given by 
${\mathcal{\hat{F}}[\hat{\varrho}]=\sum\limits_{\alpha}\hat{L}_{\alpha}\hat{\varrho}\hat{L}_{\alpha}^{\dag}}$ and ${\mathcal{\hat{F}}[\hat{\varrho}_{\mathrm{post}}]=\sum\limits_{\alpha}\hat{L}_{\alpha}\hat{\varrho}_{\mathrm{post}}\hat{L}_{\alpha}^{\dag}}$, respectively. 
The trace preserving maps are described through the Kraus operators $\hat{L}_{\alpha}$, $L_{1}=\sqrt{\mu_{1}}|1\rangle\langle 0|_{1}\otimes|0\rangle\langle 1|_{2}$ and $L_{2}=\sqrt{\mu_{2}}|0\rangle\langle 1|_{1}\otimes|1\rangle\langle 0|_{2}$, with coefficients preserving the normalization condition $\mu_{1}+\mu_{2}=1$. We define QW as follows:
\begin{eqnarray}\label{QWone}
\mathcal{W}_{\hat{\varrho}}(j,\alpha)=\big|\mathcal{P}^{j\alpha}_{\hat{\varrho}}-\mathcal{Q}^{j\alpha}_{\hat{\varrho}}\big|.
\end{eqnarray}
Here, $\mathcal{P}^{j\alpha}_{\hat{\varrho}}$ corresponds to the direct and $\mathcal{Q}^{j\alpha}_{\hat{\varrho}}$ to the indirect measurement, respectively.
After tracing out the Majorana states, both the reduced density matrix of QDs in the ground state and post-measurement density matrix read:
\begin{eqnarray}\label{densitymatrixappendix}
&&\hat\varrho=\eta_+^2\ket{\Phi_d^{-}}\bra{\Phi_d^{-}}+\xi_+^2\ket{\Psi_d^{-}}\bra{\Psi_d^{-}},\nonumber\\
&&\hat\varrho_{\text{post}}=\frac{\left(\ket{0}\bra{0}_1\otimes \mathcal{I}_2\right)\hat\varrho\left(\ket{0}\bra{0}_1\otimes \mathcal{I}_2\right)}{\text{Tr}\left[ \left(\ket{0}\bra{0}_1\otimes \mathcal{I}_2\right)\hat\varrho\left(\ket{0}\bra{0}_1\otimes \mathcal{I}_2\right)\right] },
\end{eqnarray}
where $\mathcal{I}_2$ is the identity operator acting in the spin subspace of the second QD.
Taking into account Eqns.\ (\ref{QWone},\ref{densitymatrixappendix}), and symmetric map $\mu_{1}=\mu_{2}=1/2$ we obtain after lengthy calculations 
\begin{eqnarray}\label{wittnesvalue}
\mathcal{W}_{\hat{\varrho}}(j,\alpha)=\frac{1}{4}\cdot\frac{(\omega+\Delta)^2}{4\lambda^2+(\omega+\Delta)^2}.
\end{eqnarray}
From Eq.(\ref{wittnesvalue}), we see that in the strong overlap case  ${\varepsilon_M=2\omega\gg\lambda}$, the results of  measurements in the cases (a) and (b) for the second spin may differ up to $\mathcal{W}_{\hat{\varrho}}(j,\alpha)=1/4$.

\section{Uncertainty in tripartite system}
\label{sec: tripartite system}

When quantum memory consists of two quantum systems, the lower bound of measurement uncertainties is given by the tripartite quantum-memory-assisted entropic uncertainty relation.  It is precisely our case. Therefore, to analyze the quantum memory of Majorana quasiparticles, we exploit quantum memory-assisted improved EUR for the tripartite system \cite{ming2020improved,PhysRevA.102.052227,PhysRevA.106.062219}.

Our tripartite system consists of two QDs shared by {\bf Alice} (\textbf{A}) and {\bf Bob} (\textbf{B}) and Majorana fermions owned by {\bf Charlie} (\textbf{C}): $\hat\rho_{ABC}$. {\bf Alice} measures $Z$ and $X$ components of the spin of the first QD. {\bf Bob}'s task is to minimize uncertainty of the measurement of  $X$. {\bf Charlie} tries to minimize the uncertainty of $Z$.
The quantum memory assisted EUR reads  \cite{PhysRevA.102.052227}:
\begin{equation}\label{improved tripartite first}
 S(X\vert B)+S(Z\vert C) \geqslant \log_2\frac{1}{c}  +\text{max}\lbrace 0,\delta\rbrace  
+\frac{S(A\vert B)+S(A\vert C)}{2}.
\end{equation} 
Here, the measure of complementarity is $ \log_2(1/c)=1$, $\hat\rho_{AB}=\text{Tr}_C(\hat\rho_{ABC})$,   $\hat\rho_{AC}=\text{Tr}_B(\hat\rho_{ABC})$,      $S(X\vert B)=S(\hat\rho^X_{AB})-S(\hat\rho^X_B)$, $S(Z\vert C)=S(\hat\rho^Z_{AC})-S(\hat\rho^Z_C)$, $S(A\vert B)=S(\hat\rho_{AB})-S(\hat\rho_B)$,  $S(A\vert C)=S(\hat\rho_{AC})-S(\hat\rho_C)$ are conditional quantum entropies of the states:
\begin{eqnarray}\label{afterZ}
&&\hat\rho^Z_{AC}=\sum\limits_{n=1,2}\Pi_{n}^{zA}\otimes \text{Tr}_A\left\lbrace(\Pi_{n}^{zA}\otimes\hat I_C)\text{Tr}_B\left(\hat\rho_{ABC}\right)\right\rbrace,\\
\label{afterX}
&&\hat\rho^X_{AB}=\sum\limits_{n=1,2}\Pi_{n}^{xA}\otimes \text{Tr}_A\left\lbrace(\Pi_{n}^{xA}\otimes\hat I_B)\text{Tr}_C\left(\hat\rho_{ABC}\right)\right\rbrace. 
\;\;\;\;\;
\end{eqnarray}
The second term in Eq.~(\ref{improved tripartite first}) has the form:
\begin{eqnarray}\label{Holevo first}
\delta=\frac{{I}(A:B)+{I}(A:C)}{2}-
\left[{H}(X:B)+{H}(Z:C)\right]. 
\;\;\;\;\;
\end{eqnarray}
The first two terms in Eq.~(\ref{Holevo first}) correspond to mutual quantum information 
\begin{eqnarray}
{I}(A:B)&=&S(\hat\rho_A)+S(\hat\rho_B)-S(\hat\rho_{AB}), \\
{I}(A:C)&=&S(\hat\rho_A)+S(\hat\rho_C)-S(\hat\rho_{AC}), 
\end{eqnarray}
and the second term quantifies the so-called Holevo's quantity, that is the upper bound of the information accessible to {\bf Bob} about the outcomes of {\bf Alice}'s measurements:
\begin{eqnarray}\label{Holevo second}
&&{H}(X:B)=S(\hat\rho_B)-\sum\limits_{i=1,2}p^x_iS(\hat\rho_{B|i}),
\end{eqnarray}
where
\begin{eqnarray}\label{Holevo second plus}
&&p^x_i={\rm Tr}_{AB}\left(\Pi_i^{xA}\hat\rho_{AB}\Pi_i^{xA}\right),\nonumber\\
&&\hat\rho_{B|i}=\frac{\text{Tr}_A\left(\Pi_i^{xA}\hat\rho_{AB}\Pi_i^{xA}\right)}{p^x_i},
\end{eqnarray}
and
\begin{eqnarray}\label{Holevo third}
&&
{H}(Z:C)=S(\hat\rho_C)-\sum\limits_{i=1,2}p^z_iS(\hat\rho_{C|i}),
\end{eqnarray}
where
\begin{eqnarray}\label{Holevo third plus}
&&p^z_i={\rm Tr}_{AC}\left(\Pi_i^{zA}\hat\rho_{AC}\Pi_i^{zA}\right),\nonumber\\
&&\hat\rho_{C|i}=\frac{\text{Tr}_A\left(\Pi_i^{zA}\hat\rho_{AC}\Pi_i^{zA}\right)}{p^z_i}.
\end{eqnarray}
Taking into account that $\hat\rho_{ABC}=\ket{e_1}\bra{e_1}$  after cumbersome calculations,
we deduce the expressions of the marginal entropies:


\begin{figure}
\includegraphics[width=7.5cm]{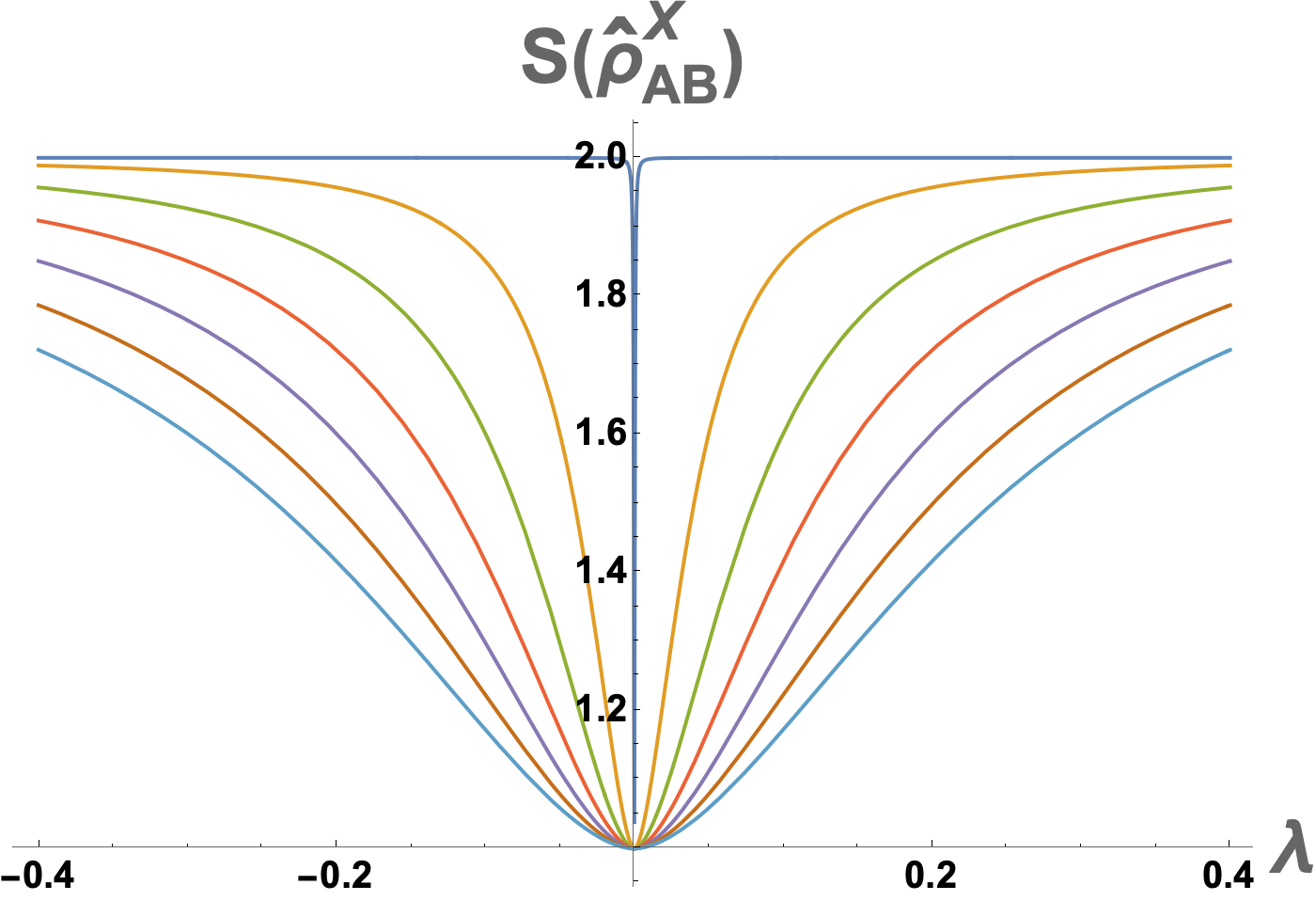}
\includegraphics[width=7.5cm]{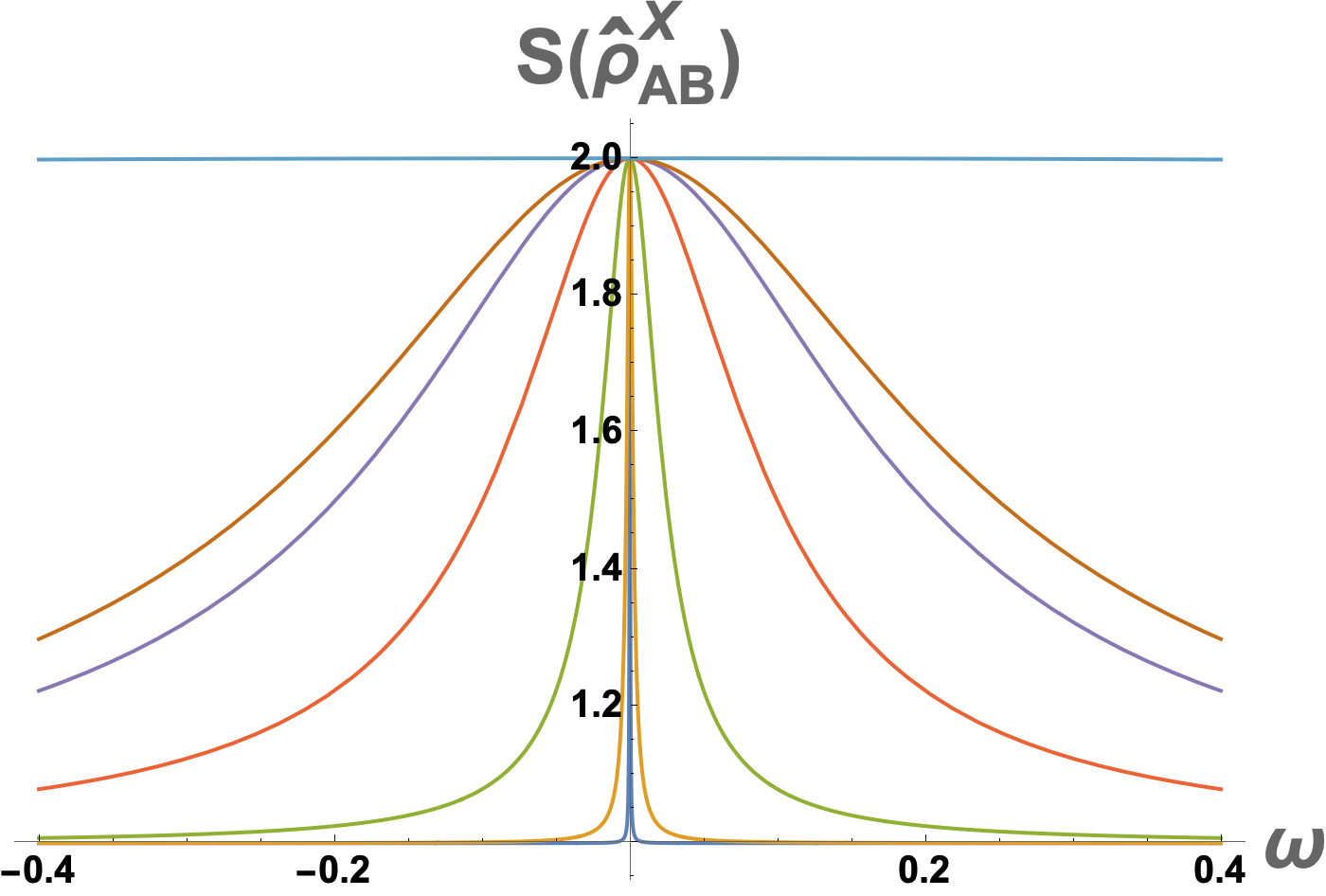}
\caption{
Entropy $S(\hat\rho^X_{AB})$ from Eq.(\ref{rhoXAB}). 
{\bf Top:} As function of the coupling strength $\lambda$ for fixed overlap between Majorana modes $\omega$ (from above $\omega=10^{-3}$, 0.1, 0.2, 0.3, 0.4, 0.5, 0.6); 
{\bf Bottom:} As function of $\omega$ for fixed $\lambda$ (from above 1, 0.1, $8\cdot 10^{-2}$, $4\cdot10^{-2}$, $10^{-2}$, $10^{-3}$, $10^{-4}$).
The entropy approaches the maximal value, as one of the limits of Eq.(\ref{eq:UpBound}) is approached. 
}
\label{fig:Sabx}
\end{figure}

\begin{eqnarray}\label{marginal entropiesAB}
&&\hat\rho_{AB}=\frac{\eta^2_+}{2}\bigg(\ket{0}\bra{0}_A\otimes\ket{0}\bra{0}_B+
\ket{1}\bra{1}_A\otimes\ket{1}\bra{1}_B-\nonumber\\
&&\ket{1}\bra{0}_A\otimes\ket{1}\bra{0}_B-
\ket{0}\bra{1}_A\otimes\ket{0}\bra{1}_B\bigg)+\nonumber\\
&&\frac{\xi^2_+}{2}\bigg(\ket{0}\bra{0}_A\otimes\ket{1}\bra{1}_B+
\ket{1}\bra{1}_A\otimes\ket{0}\bra{0}_B+\nonumber\\
&&\ket{0}\bra{1}_A\otimes\ket{1}\bra{0}_B+
\ket{1}\bra{0}_A\otimes\ket{0}\bra{1}_B\bigg),
\end{eqnarray}
and
\begin{eqnarray}\label{marginal entropiesAC}
&&\hat\rho_{AC}=\frac{\eta^2_+}{2}\ket{1}\bra{1}_C\otimes\hat I_A+\frac{\xi^2_+}{2}\ket{0}\bra{0}_C\otimes\hat I_A+\nonumber\\
&&\frac{\eta_+\xi_+}{2}(\ket{0}\bra{1}_A-\ket{1}\bra{0}_A)\otimes(\ket{0}\bra{1}_C-\ket{1}\bra{0}_C).
\end{eqnarray}
Consequently: 
\begin{eqnarray}\label{marginal entropiesA-B-C}
\hat\rho_{C}&=&\eta^2_+\ket{1}\bra{1}_C+\xi^2_+\ket{0}\bra{0}_C,\nonumber\\
\hat\rho_A&=&\frac{1}{2}\hat I_A, \hspace{0.5cm}
\hat\rho_B=\frac{1}{2}\hat I_B,
\end{eqnarray}
where $\hat I$ are identity matrices in the corresponding subspaces. We proceed with the post-measurement states, obtaining

\begin{eqnarray}\label{ZAC}
\hat\rho^Z_{AC}=\frac{1}{2}\left(\eta_+^2\ket{1}\bra{1}_C+\xi_+^2\ket{0}\bra{0}_C\right)\otimes I_A,
\end{eqnarray}
and
\begin{eqnarray}\label{XAB}
\hat\rho^X_{AB}&=&\frac{1}{4}I_AI_B+\frac{\xi^2_+-\eta_+^2}{4}\bigg(\ket{0}\bra{1}_A\otimes\ket{1}\bra{0}_B\nonumber\\
&+&\ket{0}\bra{1}_A\otimes\ket{0}\bra{1}_B+
\ket{1}\bra{0}_A\otimes\ket{1}\bra{0}_B\nonumber\\
&+&\ket{1}\bra{0}_A\otimes\ket{0}\bra{1}_B\bigg).
\end{eqnarray}
On the other hand, for the Holevo's entropies, we obtain:
\begin{eqnarray}\label{For Holevo's quantities1}
&&\hat\rho_{C|1}=\hat\rho_{C|2}=\eta^2_+\ket{1}\bra{1}_C+
\xi^2_+\ket{1}\bra{1}_C,\nonumber\\
&&p^z_1=p^z_2=\frac{1}{2}, 
\end{eqnarray}
and
\begin{eqnarray}\label{For Holevo's quantities2}
&&\hat\rho_{B|1,2}=\frac{1}{2}I_B\pm\frac{\xi^2_+-\eta^2_+}{2}\left(\ket{0}\bra{1}_B+\ket{1}\bra{0}_B\right),\nonumber\\
&&p^x_1=p^x_2=\frac{1}{2}. 
\end{eqnarray}
Taking into account Eqs.~(\ref{afterZ})-(\ref{For Holevo's quantities2}),
we deduce the entropy of the state $\hat\rho^X_{AB}$:
\begin{eqnarray}\label{rhoXAB}
S(\hat\rho^X_{AB})&=&-\left(\frac{1-|\xi^2_+-\eta^2_+|}{2}\right)\log_2\left(\frac{1-|\xi^2_+-\eta^2_+|}{4}\right)\nonumber\\
&&-\left(\frac{1+|\xi^2_+-\eta^2_+|}{2}\right)\log_2\left(\frac{1+|\xi^2_+-\eta^2_+|}{4}\right).
\;\;\;\;\;\;
\end{eqnarray}
The entropy  $S(\hat\rho^X_{AB})$ approaches a universal upper bound for $\lambda/\varepsilon_M\gg 1$
\begin{equation}
\label{eq:UpBound}
S(\hat\rho^X_{AB}) \to 2  ,
\end{equation}
which is the entropy of the maximally entangled state on the space of hermitian $4\times 4$ matrices. 
We plot the function $S(\hat\rho^X_{AB})$ for different cases in Fig.~\ref{fig:Sabx}.
The conditional quantum entropy corresponding to the state $\hat\rho^X_{AB}$ is:
\begin{eqnarray}\label{final1}
&&S(X\vert B) = S(\hat\rho^X_{AB})-1.
\end{eqnarray}
In a similar way, we deduce
\begin{eqnarray}
\label{rhoZAC}
S(\hat\rho^Z_{AC})&=&-\xi^2_+\log_2\frac{\xi^2_+}{2}-\eta^2_+\log_2\frac{\eta^2_+}{2},\\
\label{final2}
S(Z\vert C)&=&-\xi^2_+\log_2\frac{\xi^2_+}{2}-\eta^2_+\log_2\frac{\eta^2_+}{2}\nonumber\\
&&+\xi_+^2\log_2\xi_+^2+\eta_+^2\log_2\eta_+^2=1,
\\
S(A\vert B)&=& -\xi^2_+\log_2\xi^2_+ - \eta^2_+\log_2\eta^2_+-1,
\label{final3}
\end{eqnarray}
and 
\begin{eqnarray}\label{rhoAC}
&&S(\hat\rho_{AC})=1,
\end{eqnarray}
while
\begin{eqnarray}\label{rhoAB}
S(\hat\rho_{AB})&=&-\xi_+^2\log_2\xi_+^2-\eta_+^2\log_2\eta_+^2,\\
\label{final4}
S(A\vert C)&=&1+\xi_+^2\log_2\xi_+^2+\eta_+^2\log_2\eta_+^2.
\end{eqnarray}
In the limit of $\lambda/\varepsilon_M\gg 1$, the entropy approaches the bound $S(\hat\rho_{AB})\to1$, which is exactly half the value of $S(\hat\rho^X_{AB})$.
Except this different normalization, the overall shape of both quantities is very similar. For the Holevo's terms we deduce the mutual quantum entropies:
\begin{eqnarray}
\label{final5}
{I}(A:B) &=& 2+\xi^2_+\log_2\xi^2_++\eta^2_+\log_2\eta^2_+,\\
\label{final6}
{I}(A:C) &=& -\xi^2_+\log_2\xi^2_+-\eta^2_+\log_2\eta^2_+,
\end{eqnarray}
as well as the Holevo quantity itself: 
\begin{eqnarray}
\label{final7}
{H}(X:B)=1 &+& \frac{1-\xi^2_++\eta^2_+}{2}\log_2\left(\frac{1-\xi^2_++\eta^2_+}{2}\right)\nonumber\\
&+&\frac{1+\xi^2_+-\eta^2_+}{2}\log_2\left(\frac{1+\xi^2_+-\eta^2_+}{2}\right). \;\;\;\;
\end{eqnarray}
\begin{eqnarray}\label{final8}
&&\mathcal{H}(Z:C)=0.
\end{eqnarray}
We plot the function ${H}(X:B)$ in Fig.~\ref{fig:Hxb}. 
Thus, we finally obtain:
\begin{eqnarray}
\label{inequality}
\nonumber
 1+\text{max}(0,\delta) &\leqslant&
-\left(\frac{1-|\mathcal{A}|}{2}\right)\log_2\left(\frac{1-|\mathcal{A}|}{4}\right)\\
&&-\left(\frac{1+|\mathcal{A}|}{2}\right)\log_2\left(\frac{1+|\mathcal{A}|}{4}\right),
\end{eqnarray}
where
\begin{eqnarray}
\delta=-\frac{1-\mathcal{A}}{2}\log_2\left(\frac{1-\mathcal{A}}{2}\right)
-\frac{1+\mathcal{A}}{2}\log_2\left(\frac{1+\mathcal{A}}{2}\right),\;\;\;\;
\end{eqnarray}
and $\mathcal{A}=\xi^2_+-\eta^2_+$.
Notice that for any $\mathcal{A}$, Eq.~(\ref{inequality}) converts into equality $\log_22=1$, corresponding to the minimal values of measurement uncertainties. We thus proved that the topological Majorana quasiparticles enabled the uncertainties to reach their minimal lower bound value. This analytical result is universal and holds for arbitrary coupling strength and any finite overlap between the Majorana boundary modes.

\begin{figure*}
\includegraphics[width=7.5cm]{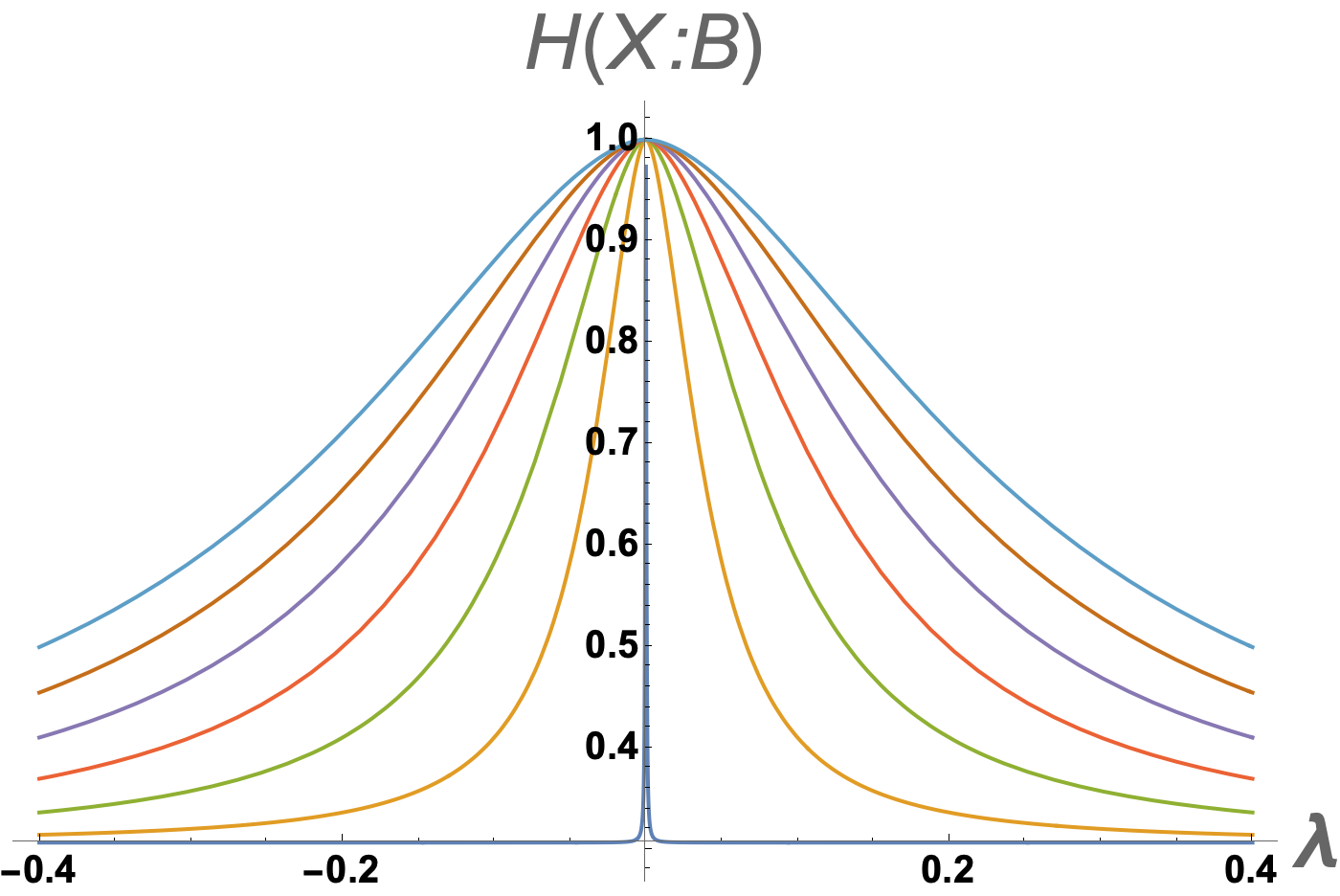}
\hspace{5mm}
\includegraphics[width=7.5cm]{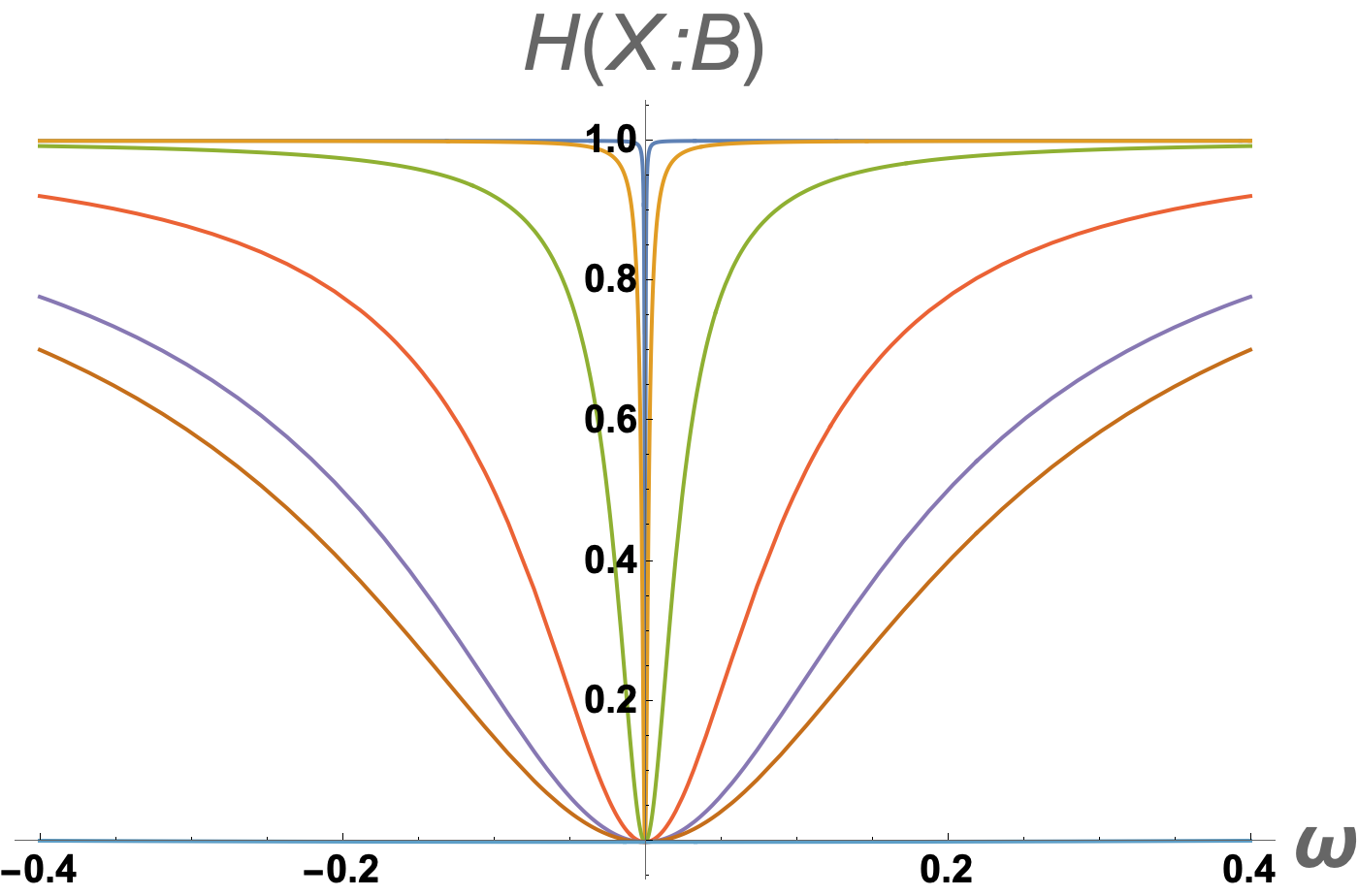}
\caption{
Holevo quantity ${H}(X:B)$ from Eq.(\ref{final7}) calculated for the same parameter sets as in Fig.~\ref{fig:Sabx}. 
{\bf Left:} As function of the coupling strength $\lambda$ for fixed overlap between Majorana modes $\omega$ (from above $\omega=10^{-3}$, 0.1, 0.2, 0.3, 0.4, 0.5, 0.6); 
{\bf Right:} As function of $\omega$ for fixed $\lambda$ (from above 1, 0.1, $8\cdot 10^{-2}$, $4\cdot10^{-2}$, $10^{-2}$, $10^{-3}$, $10^{-4}$).
}
\label{fig:Hxb}
\end{figure*}

\section{Summary and outlook}
\label{sec: Summary and outlook }

In this article we have investigated the uncertainties of POVMs performed on two quantum dots (QD$_1$, QD$_2$) interconnected through the topological superconducting nanowire, hosting the Majorana boundary modes.  
We determined the quantitative measures of the quantum dot entanglement in this tripartite system, implementing the improved entropic uncertainty relations (EUR). Moreover, we analyzed the role of mutual quantum correlations between QDs transmitted through the Majorana modes, focusing on measurements done on QD$_1$. We derived analytical expressions for the particular values of QDs energies, $\varepsilon^{}_{1\sigma}=0=\varepsilon^{}_{2\sigma}$, and the couplings with Majorana modes, $\lambda^{}_1=-\lambda^{}_2=\sqrt{2}\lambda$. Under such conditions, we have predicted that the uncertainties of measurements allowed by EUR are minimal. This result is valid for arbitrary strength of the coupling $\lambda$ and for any overlap between the Majorana modes, $\varepsilon_M\neq 0$. 
We have also observed that quantum entropies of the reduced density matrices, Eqs.~(\ref{rhoXAB}) and (\ref{rhoAB}), tend to higher values for weaker overlaps between the Majorana modes $\varepsilon_M$ and stronger coupling with QDs $\lambda$. This remarkable behavior is shown in Fig.~\ref{fig:Sabx}. According to this scenario, the interference between the Majorana modes reduces the quantum correlation between quantum dots, as respective quantum entropies are always at their maxima for $\varepsilon_M/\lambda\rightarrow 0$. This is even more surprising, since the Majorana quasiparticles themselves represent ultra-quantum fermionic excitations.
The uncertainty of measurements always coincides with the lower bound of EUR. However, its minimal value is achieved for $\mathcal{A}\rightarrow 1$, $\varepsilon_M\gg\lambda$ and scales as 
\begin{equation}
\sim 1 + \frac{1}{4\ln(2)} \frac{\lambda^2}{\varepsilon^2_M}\left( 1 + \ln\left[\frac{4\varepsilon^2_M}{\lambda^2}\right] \right),
\end{equation}
while the maximal value of uncertainty 
\begin{equation}
\sim 2 - \frac{1}{32 \ln(2)} \frac{\varepsilon^2_M}{\lambda^2}
\end{equation}
corresponds to the case  $\mathcal{A}\rightarrow 0$, $\varepsilon_M\ll\lambda$. It is a surprising and important finding since, after rigorous mathematical discussion, the main result is expressed in terms of two fundamental parameters, such as the overlap of Majorana modes $\varepsilon_M$ and the coupling strength with QDs $\lambda$. 

We note that of vital importance for the obtained results are the eigenstates of the system. Since we are interested in perspectives of a long-distance entanglement, the Majorana modes would be essential for entangling the two quantum dots communicating via topological superconductor. Coupling QDs through the Majorana modes allows for the minimization of environmental effects, reducing decoherence in long-distance communication.

Empirical verification of the multipartite entanglement in bulk materials is usually done using neutron scattering \cite{PhysRevResearch.2.043329,PhysRevB.103.224434}. These techniques have been successfully adopted for probing quantum entanglement in quantum magnets \cite{PhysRevLett.127.037201}, strongly correlated fermions \cite{PhysRevLett.130.106902,PhysRevB.106.085110},
electronic orbitals \cite{ren2024} and numerous other systems \cite{liu2024,SCHEIE2025100020,https://doi.org/10.1002/qute.202400196}.
 Such techniques, however, would be hard to use for the setup with quantum dots because the quasiparticle states are too fragile for probing via crystal spectroscopy.
Implementation of these techniques for the present setup might be also problematic due to detrimental effects on electron pairing. More promising method can rely on time-resolved charge transport measurements, where the crossed Andreev reflections could detect mutual correlations between the quantum dots \cite{PhysRevB.110.035413,teleportation-2016,teleportation-2024}. 

Further extension of our study could explore correlations, originating from the Coulomb repulsion between opposite-spin electrons at each quantum dot. Determination of the quantum entanglement of the strongly correlated dimers coupled even to non-topological environment would require, however, sophisticated numerical tools \cite{Yoshioka-2025}. Analytical results would be hardly feasible for such setup,
hence we leave this topic for further studies.

{\it Acknowledgements.} 
I.W. and T.D. acknowledge support by the National Science Centre (Poland) through the grant No.\ 2022/04/Y/ST3/00061.
A.S. acknowledges funding by the grant JSF-22-10-0001 of the Julian Schwinger Foundation for Physics Research.


\bibliography{memory.bib}

\end{document}